\documentclass[conference]{IEEEtran}
\usepackage{amsmath}
\usepackage{amssymb}
\usepackage{amsfonts}
\usepackage{amscd}
\usepackage{mathrsfs}
\usepackage{graphicx}
\usepackage{color}
\usepackage{url}
\usepackage{bm}
\usepackage{algorithm}
\usepackage{algorithmic}
\usepackage{setspace}

\newtheorem{theorem}{Theorem}
\newtheorem{lemma}{Lemma}
\newtheorem{definition}{Definition}
\newtheorem{proposition}{Proposition}

\newcommand{\samp}{\hookleftarrow}

\newcommand{\mR}{\mathbb{R}}

\newcommand{\poly}{\mathrm{poly}}
\newcommand{\eps}{\varepsilon}

\newcommand{\aln}[1]{\ensuremath{\mathsf{#1}}}
\IEEEoverridecommandlockouts

\begin{document}
\title{On Massive \aln{MIMO} Physical Layer Cryptosystem}
\author{\IEEEauthorblockN{Ron Steinfeld and Amin Sakzad}
\IEEEauthorblockA{Clayton School of Information Technology\\
Monash University, Melbourne,Victoria, Australia\\
Emails: {\tt ron.steinfeld and amin.sakzad@monash.edu}}}
\maketitle
\begin{abstract}
In this paper, we present a zero-forcing (\aln{ZF}) attack on the physical layer cryptography scheme based on massive multiple-input multiple-output (\aln{MIMO}). The scheme uses singular value decomposition (\aln{SVD}) precoder. We show that the eavesdropper can decrypt/decode the information data under the same condition as the legitimate receiver. We then study the advantage for decoding by the legitimate user over the eavesdropper in a generalized scheme using an arbitrary precoder at the transmitter. On the negative side, we show that if the eavesdropper uses a number of receive antennas much larger than the number of legitimate user antennas, then there is no advantage, independent of the precoding scheme employed at the transmitter. On the positive side, for the case where the adversary is limited to have the same number of antennas as legitimate users, we give an $\mathcal{O}\left(n^2\right)$ upper bound on the advantage and show that this bound can be approached using an inverse precoder.
\end{abstract}
\section{Introduction}

Recently, an interesting new approach for physical security in massive multiple-input multiple-output (\aln{MIMO}) communication systems was introduced by Dean and Goldsmith~\cite{DG13} and called ``Physical layer cryptography'', or a massive \aln{MIMO} physical layer cryptosystem (\aln{MM-PLC}). In this scenario, the channel state information (\aln{CSI}) is known at the legitimate transmitter as well as all the other adversaries and legitimate receivers. The eavesdropper has also the knowledge of the \aln{CSI} between legitimate users. The idea is to replace the information-theoretic security guarantees of previous physical layer security methods 
with the weaker complexity-based security guarantees used in cryptography. More precisely, the idea of~\cite{DG13} is to precode the information data at the transmitter, based on the known \aln{CSI} between the legitimate users, so that the decoding of the received vector would be computationally easy for the legitimate user but computationally hard for the adversary. The goal of this approach is to trade-off a weaker, but still practical, complexity-based security guarantee in order to avoid the less practical additional assumptions required by existing information-theoretic techniques, such as higher noise level in~\cite{OH11,ZSB13,WLWQ14} and/or less antennas for the adversary than for legitimate parties in~\cite{GN08}, while still retaining the ``no secret key'' location-based decryption feature of physical-layer security methods.


In~\cite{DG13}, a \aln{MM-PLC} is presented that is claimed to achieve the above goal of the complexity-based approach, using a singular value decomposition (\aln{SVD}) precoding technique and $m$-PAM constellations at the transmitter. Namely, it is claimed that, under a certain condition on the number $n_t$ of legitimate sender's transmit antennas and the noise level $\beta$ in the adversary's channel (which we call the \emph{hardness condition} of~\cite{DG13}), the message decoding problem for the adversary (eavesdropper), termed the \aln{MIMO-Search} problem in~\cite{DG13}, is as hard to solve on average as it is to solve a standard conjectured hard lattice problem in dimension $n_t$ in the worst-case, in particular, the $\aln{GapSVP}_{\poly(n_t)}$ variant of the approximate shortest vector problem in arbitrary lattices of dimension $n_t$, with approximation factor polynomial in $n_t$. For these problems, no polynomial-time algorithm is known, and the best known algorithms run in time exponential in the number of transmit antennas $n_t$, which is typically infeasible when $n_t$ is in the range of few hundreds (as in the case of massive \aln{MIMO}). Significantly, this computational hardness of \aln{MIMO-Search} is claimed to hold even if the adversary is allowed to use a large number of receive antennas $n'_r = \poly(n_t)$ \emph{polynomially larger} than $n_t$ and $n_r$ used by the legitimate parties, and with the same noise level as the legitimate receiver ($\beta=\alpha$). Consequently, under the widely believed conjecture that no polynomial-time algorithms for $\aln{GapSVP}_{\poly(n_t)}$ in dimension $n_t$ exist and the hardness condition of~\cite{DG13}, the authors of~\cite{DG13} conclude that their \aln{MM-PLC} and the corresponding \aln{MIMO-Search} problem is secure against adversaries with run-time polynomial in $n_t$.

\emph{Our Contribution.} In this contribution, we further analyse the complexity-based \aln{MM-PLC} initiated in~\cite{DG13}, to improve the understanding of its potential and limitations. Our contributions are summarized below:
\begin{itemize}
\item We show, using a linear receiver known as zero-forcing (\aln{ZF})~\cite{Kumar09}, an algorithm with run-time polynomial in $n_t$ for the \aln{MIMO-Search} problem faced by an adversary against the \aln{MM-PLC} in~\cite{DG13}. We analyze the decoding success probability of this algorithm and prove that it is $\geq 1-o(1)$ even if the \emph{hardness condition} of~\cite{DG13} is satisfied, if the ratio $y'=n_r'/n_t$ exceeds a small factor at most logarithmic in $n_t$, i.e. $y' = \mathcal{O}(\log n_t)$. This contradicts the hardness of the \aln{MIMO-Search} problem conjectured in~\cite{DG13} to hold for much larger polynomial ratios $y' =\mathcal{O}(\poly(n_t))$.  Moreover, we show that the decoding success probability of an adversary against the \aln{MM-PLC} of~\cite{DG13} using the \aln{ZF} decoder is approximately the same (or greater than) as the decoding success probability of the legitimate receiver if $n'_r$ is approximately greater than or equal to $n_r$, assuming an equal noise level for adversary and legitimate receivers. Our first contribution implies that the \aln{SVD} precoder-based \aln{MM-PLC} in~\cite{DG13} still requires for security an undesirable assumption limiting $n'_r$ to be less than that of the legitimate receiver, similar to previous information-theoretic techniques.

\item As our second contribution, we investigate the potential of the general approach of~\cite{DG13} assuming \aln{ZF} decoding by the both adversary and legitimate receiver, by studying the generalized scenario where one allows arbitrary precoding matrices by the legitimate transmitter in place of the \aln{SVD} precoder of the scheme in~\cite{DG13}. To do so, we define a decoding advantage ratio for the legitimate user over the adversary, which is approximately the ratio of the maximum noise power tolerated by the legitimate user's decoder to the maximum noise power tolerated by the adversary's decoder (for the same ``high'' success probability). We derive a general upper bound on this advantage ratio, and show that, even in the general scenario, the advantage ratio tends to 1 (implying no advantage), if the ratio $n'_r/\max(n_t,n_r)$ exceeds a small constant factor ($\leq 9$). Thus a linear limitation (in the number of legitimate user antennas) on the number of adversary antennas seems inherent to the security of this approach. On the positive side, we show that, in the case when legitimate parties and the adversary all have the same number of antennas ($n'_r=n_r=n_t$), the upper bound on the advantage ratio is quadratic in $n_t$ and we give experimental evidence that this upper bound can be approximately achieved using an inverse precoder.
\end{itemize}
{\bf Notation.} The notation $a\gg b$ denotes that the real number $a$ is much greater than $b$. We let $|z|$ denotes the absolute value of $z$. Vectors will be column-wise and denoted by bold small letters. Let ${\bf v}$ be a vector, then its $j$-th entry is represented by $v_j$. A $k_1\times k_2$ matrix ${\bf X}=[{\bf x}_1,\ldots,{\bf x}_{k_2}]$ is formed by joining the $k_1$-dimensional column vectors ${\bf x}_1,\ldots, {\bf x}_{k_2}$. The superscript $^t$ denotes transposition operation. We make use of the standard Landau notations to classify the growth of functions. We say that a function $F(n)$ is $\poly(n)$ if it is bounded by a polynomial in $n$. The notation $\omega(F(n))$ refers to the set of functions (or an arbitrary function in that set) growing faster than $cF(n)$ for any constant $c > 0$. A function $G(n)$ is said negligible if it is proportional to $n^{-\omega(1)}$. If $X$ is a random variable, $\mathbb{P}[X = x]$ denotes the probability of the event ``$X = x$''. The standard Gaussian distribution on $\mR$ with zero mean and variance $\sigma^2$ is denoted by $\mathcal{N}_{\sigma^2}$. We denote by $w \samp \mathcal{D}$ the assignment to random variable $w$ a sample from the probability distribution $\mathcal{D}$.
\section{System Model}\label{Section:Background}
We first summarize the notion of real lattices and \aln{SVD} (of a matrix) which are essential for the rest of the paper. A $k$-dimensional {\em lattice} $\Lambda$ with a basis set $\{{\boldsymbol\ell}_1,\ldots,{\boldsymbol\ell}_k\}\subseteq\mathbb{R}^d$ is the set of all integer linear combinations of basis vectors.
Every matrix ${\bf M}_{s\times t}$ admits a singular value decomposition (\aln{SVD}) ${\bf M}={\bf U}{\bf \Sigma}{\bf V}^t$, where the matrices ${\bf U}_{s\times s}$ and ${\bf V}_{t\times t}$ are two orthogonal matrices and ${\bf \Sigma}_{s\times t}$ is a rectangular diagonal matrix with non-negative diagonal elements $\sigma_1({\bf M})\geq \cdots\geq\sigma_s({\bf M})$. By abusing the notation, we denote the Moore--Penrose pseudo-inverse of ${\bf M}$ by ${\bf M}^{-1}$, that is ${\bf V}{\bf \Sigma}^{-1}{\bf U}^t$, where the pseudo-inverse of ${\bf \Sigma}$ is denoted by ${\bf \Sigma}^{-1}$ and can be obtained by taking the reciprocal of each non-zero entry on the diagonal of ${\bf \Sigma}$ and finally transposing the matrix.
\subsection{Dean-Goldsmith Model}\label{SubSection:SystemModel}
We consider a slow-fading \aln{MIMO} wiretap channel model. 
The $n_r\times n_t$ real-valued \aln{MIMO} channel from user $\mathrm{A}$ to user $\mathrm{B}$ is denoted by ${\bf H}$. We also denote the channel from $\mathrm{A}$ to the adversary $\mathrm{E}$ by an $n_r'\times n_t$ matrix ${\bf G}$. The entries of ${\bf H}$ and ${\bf G}$ are identically and independently distributed (i.i.d.) based on a Gaussian distribution $\mathcal{N}_1$. These channel matrices are assumed to be constant for long time as we employ precoders at the transmitter. This model can be written as:
$$\left\{\begin{array}{l}
{\bf y} = {\bf H}{\bf x} + {\bf e},\\
{\bf y}' = {\bf G}{\bf x} + {\bf e}'.
\end{array}\right.$$
The entries $x_i$ of ${\bf x} \in \mathbb{R}^{n_t}$, for $1\leq i \leq n_t$, are drawn from a constellation $\mathcal{X}=\{0,1,\ldots, m-1\}$ for an integer $m$.
The components of the noise vectors ${\bf e}$ and ${\bf e}'$ are i.i.d. based on Gaussian distributions $\mathcal{N}_{m^2\alpha^2}$ and $\mathcal{N}_{m^2\beta^2}$, respectively. We assume $\alpha = \beta$ to evaluate the potential of the Dean-Goldsmith model to provide security based on computational complexity assumptions, without a ``degraded noise'' assumption on the eavesdropper.
In this communication setup, the \aln{CSI} is available at all the transmitter and receivers. In fact, users $\mathrm{A}$ and $\mathrm{B}$ know the channel matrix ${\bf H}$ (via some channel identification process), while adversary $\mathrm{E}$ has the knowledge of both channel matrices ${\bf G}$ and ${\bf H}$. The knowledge of ${\bf H}$ allows $\mathrm{A}$ to perform a linear precoding to the message before transmission. More specifically, in~\cite{DG13}, to send a message ${\bf x}$ to $\mathrm{B}$, user $\mathrm{A}$ performs an \aln{SVD} precoding as follows. Let \aln{SVD} of ${\bf H}$ be given as ${\bf H} = {\bf U}{\bf \Sigma}{\bf V}^t$. The user $\mathrm{A}$ transmits ${\bf V}{\bf x}$ instead of ${\bf x}$ and $\mathrm{B}$ applies a filter matrix ${\bf U}^t$ to the received vector ${\bf y}$. With this, the received vectors at $\mathrm{B}$ and $\mathrm{E}$ are as follows:
$$\left\{\begin{array}{l}
\tilde{\bf y} = {\bf \Sigma}{\bf x} + \tilde{\bf e},\\
{\bf y}' = {\bf G}{\bf V}{\bf x} + {\bf e}',
\end{array}\right.$$
where $\tilde{\bf e}={\bf U}^t{\bf e}$. Note that since ${\bf U}^t$ and ${\bf V}$ are both orthogonal matrices, the vector $\tilde{\bf e}$ and the matrix ${\bf G}_v\triangleq{\bf G}{\bf V}$ continue to be i.i.d. Gaussian vector and matrix, with components of zero mean and variances $m^2\alpha^2$ and $1$, respectively.
\subsection{Correctness Condition}\label{SubSection:CorrectnessCondition}
Although Dean-Goldsmith do not provide a correctness analysis, we provide one here for completeness. Since ${\bf \Sigma} = \mbox{diag}(\sigma_1({\bf H}),\ldots,\sigma_{n_t}({\bf H}))$ is diagonal, user $\mathrm{B}$ recovers an estimate $\tilde{x}_i$ of the $i$-th coordinate/layer $x_i$ of ${\bf x}$, by performing two operations dividing and rounding as follows:
$\tilde{x}_i = \left\lceil \tilde{y}_i/\sigma_i({\bf H})\right\rfloor = x_i + \left\lceil \tilde{e}_i/\sigma_i({\bf H})\right\rfloor$.
It is now easy to see that the decoding process succeeds if $|\tilde{e}_i| < |\sigma_i({\bf H})|/2$ for all $1\leq i\leq n_t$.
Since each $\tilde{e}_i$ is distributed as $\mathcal{N}_{m^2\alpha^2}$, the decoding error probability, $\mathbb{P}(\mathrm{B}|{\bf H})$ that $\mathrm{B}$ incorrectly decodes ${\bf x}$, is, by a union bound, upper bounded by $n_t$ times the probability of decoding error at the worst layer:
\begin{eqnarray}
\mathbb{P}(\mathrm{B}|{\bf H})\!\!\!\!&\leq&\!\!\!\! n_t\mathbb{P}_{w \samp \mathcal{N}_{m^2 \alpha^2}}\left(|w| < |\sigma_{n_t}({\bf H})|/2\right)\nonumber\\
\!\!\!\!&=&\!\!\!\!n_t\mathbb{P}_{w \samp \mathcal{N}_{1}}\left(|w| < |\sigma_{n_t}({\bf H})|/(2 m \alpha) \right) ~\label{Pe:AB}\\
\!\!\!\!&\leq&\!\!\!\!n_t \exp\left(-|\sigma_{n_t}({\bf H})|^2/(8 m^2 \alpha^2)\right)\!,\nonumber
\end{eqnarray}
where we have used the bound $\exp(-x^2/2)$ on the tail of the standard Gaussian distribution. By choosing parameters such that $m^2 \alpha^2 \leq |\sigma_{n_t}({\bf H})|^2/(8 \log(n_t/\eps))$, one can ensure that $\mathrm{B}$'s error probability $\mathbb{P}(\mathrm{B}|{\bf H})$ is less than any $\eps>0$.
\subsection{Security Condition}\label{SubSection:SecurityCondition}
Unlike decoding by user $\mathrm{B}$, for decoding by the adversary $\mathrm{E}$, the authors of~\cite{DG13} claimed that the complexity of a problem called in~\cite{DG13} the ``Search'' variant of the ``\aln{MIMO} decoding problem'' (to be called \aln{MIMO-Search} from here on), namely recovering ${\bf x}$ from ${\bf y}' = {\bf G}_v{\bf x} + {\bf e}'$ and ${\bf G}_v$, with non-negligible probability, under certain parameter settings, upon using massive \aln{MIMO} systems with large number of transmit antennas $n_t$, is as hard as solving standard lattice problems in the worst-case.
More precisely, it was claimed in~\cite{DG13} that, upon considering above conditions, user $\mathrm{E}$ will face an exponential complexity in decoding the message ${\bf x}$. The above cryptosystem is called the {\em Massive \aln{MIMO} Physical Layer Cryptosystem (\aln{MM-PLC})}, and the above problem of recovering ${\bf x}$ from ${\bf y}'$ is called in~\cite{DG13} the ``Search'' variant of the ``\aln{MIMO} decoding problem''. For our security analysis, we focus here for simplicity on this \aln{MIMO-Search} variant. We say that the \aln{MIMO-Search} problem is \emph{hard} (and the \aln{MM-PLC} is \emph{secure} in the sense of ``one-wayness'') if any attack algorithm against \aln{MIMO-Search} with run-time $\poly(n_t)$ has negligible success probability $n_t^{-\omega(1)}$.
More precisely, in Theorem $1$ of~\cite{DG13}, a polynomial-time complexity reduction is claimed from worst-case instances of the $\aln{GapSVP}_{n_t/\alpha}$ problem in arbitrary lattices of dimension $n_t$, to the \aln{MIMO-Search} problem with $n_t$ transmit antennas, noise parameter $\alpha$ and constellation size $m$, assuming the following minimum noise level for the equivalent channel in between $\mathrm{A}$ and $\mathrm{E}$ holds:
\begin{equation}~\label{eq:const1}
m\alpha>\sqrt{n_t}.
\end{equation}
The reduction is quantum when $m = \poly(n_t)$ and classical when $m = \mathcal{O}(2^{n_t})$, and is claimed to hold for \emph{any polynomial number of receive antennas} $n'_r = \poly(n_t)$. We show in the next Section, however, that in fact for $m \alpha < c n'_r/\sqrt{\log n_t}$ for some constant $c$, there exists an efficient algorithm for \aln{MIMO-Search}. 
Since \eqref{eq:const1} is independent of the number of receive antennas $n_r'$, the condition \eqref{eq:const1} turns out to be not sufficient to provide security of the \aln{MM-PLC}. We will provide our detailed analysis in the next Section.
\section{Zero-Forcing Attack}\label{Section:Zero-ForcingAttack}
In this section, we introduce a simple and efficient attack based on \aln{ZF} linear receivers~\cite{Kumar09}. We first introduce the attack and analyze its components. The eavesdropper $\mathrm{E}$ receives ${\bf y}' = {\bf G}_v{\bf x} + {\bf e}'$. Let ${\bf G}_v = {\bf U}'{\bf \Sigma}'({\bf V}')^t$ be the \aln{SVD} of the equivalent channel ${\bf G}_v$. Thus, we get
${\bf y}' = {\bf U}'{\bf \Sigma}'({\bf V}')^t{\bf x} + {\bf e}'$,
where both ${\bf U}'$ and ${\bf V}'$ are orthogonal matrices and ${\bf \Sigma}'$ equals
$\mbox{diag}\left(\sigma_1({\bf G}_v),\ldots,\sigma_{n_t}({\bf G}_v)\right)=\mbox{diag}\left(\sigma_1({\bf G}),\ldots,\sigma_{n_t}({\bf G})\right)$,
where the last equality holds since the singular values of ${\bf G}_v$ and ${\bf G}$ are the same. Note that $\mathrm{E}$ knows ${\bf G}_v$ and its \aln{SVD} from the assumption that (s)he knows the channel between $\mathrm{A}$ and $\mathrm{B}$. At this point, user $\mathrm{E}$ performs a \aln{ZF} attack~\cite{Kumar09}. S(he) computes
\begin{equation}~\label{eq:Eveequation}
\tilde{\bf y}' = ({\bf G}_v)^{-1} {\bf y}' = {\bf x} + \tilde{\bf e}',
\end{equation}
where $\tilde{\bf e}' = ({\bf G}_v)^{-1}{\bf e}' = {\bf V}'({\bf \Sigma}')^{-1}({\bf U'})^t{\bf e}'$. User $\mathrm{E}$ is now able to recover an estimate $\tilde{x}'_i$ of the $i$-th coordinate $x_i$ of ${\bf x}$, by rounding:
$\tilde{x}'_i = \left\lceil \tilde{y}_i'\right\rfloor = \left\lceil x_i + \tilde{e}'_i\right\rfloor = x_i + \left\lceil\tilde{e}'_i\right\rfloor$.
\subsection{Analysis of ZF Attack}\label{SubSection:NoiseAnalysis}
We now investigate the distribution of $\tilde{\bf e}'$ in \eqref{eq:Eveequation}.
\begin{lemma}~\label{lem:noiseatEVE}
The components of $\tilde{\bf e}'$ in \eqref{eq:Eveequation} are distributed as $\mathcal{N}_{\sigma_{\mathrm{E}}^2}$ with
$\sigma_{\mathrm{E}}^2\leq(m^2\alpha^2)/\sigma^2_{n_t}({\bf G})$.
\end{lemma}
\begin{IEEEproof}
Note that $({\bf U}')^t{\bf e}'$ has the same distribution as ${\bf e}'$ since $({\bf U}')^t$ is orthogonal. Hence, $z_j$, the $j$-th coordinate of the vector ${\bf z} = ({\bf \Sigma}')^{-1}({\bf U'})^t{\bf e}'$ is distributed as $\mathcal{N}_{m^2\alpha^2/\sigma^2_j({\bf G})}$, for all $1\leq j\leq n_t$. We also note that $z_j$'s are independent with different variances. Now let ${\bf v}_i'$ denotes the $i$-th row of ${\bf V}'$. We find the distribution of
\begin{equation}~\label{eq:innerproduct}
t_i = \langle{\bf v}_i',{\bf z}\rangle= \sum_{j=1}^{n_t} v'_{i,j}z_j.
\end{equation}
Since 
the linear combination at \eqref{eq:innerproduct} is distributed as a linear combination of independent Gaussian distributions, $t_i$ is distributed as
\begin{eqnarray}
\sum_{j=1}^{n_t} v'_{i,j}\mathcal{N}_{m^2\alpha^2/\sigma^2_j({\bf G})}\!\!\!\! &=&\!\!\!\! \mathcal{N}_{\sum_{j=1}^{n_t} |v'_{i,j}|^2m^2\alpha^2/\sigma^2_j({\bf G})}\label{eq:inner1}\\
\!\!\!\!&=&\!\!\!\! \mathcal{N}_{m^2\alpha^2\sum_{j=1}^{n_t} |v'_{i,j}|^2/\sigma^2_j({\bf G})}.\label{eq:inner2}
\end{eqnarray}
Since $\sigma^2_j({\bf G})\geq \sigma^2_{n_t}({\bf G})$, for all $1\leq j\leq n_t$, the random variable $t_i$ is distributed as $\mathcal{N}_{\sigma^2_{t_i}}$ with
\begin{equation}\label{eq:inn3}
\sigma^2_{t_i}\!=\!m^2\alpha^2\sum_{j=1}^{n_t} \!\frac{|v'_{i,j}|^2}{\sigma^2_j({\bf G})}\!\leq\!\frac{m^2\alpha^2}{\sigma^2_{n_t}({\bf G})}\sum_{j=1}^{n_t} |v'_{i,j}|^2=\!\frac{m^2\alpha^2}{\sigma^2_{n_t}({\bf G})},
\end{equation}
where the last equality holds because ${\bf V}'$ is orthogonal.
\end{IEEEproof}
The above explained \aln{ZF} attack succeeds if $|\tilde{e}'_i| < 1/2$ for all $1\leq i \leq n_t$. Let $\mathbb{P}_{\mbox{\tiny ZF}}(\mathrm{E}|{\bf G})$ denotes the decoding error probability that $\mathrm{E}$ incorrectly recovers ${\bf x}$ using \aln{ZF} attack. Based on Lemma~\ref{lem:noiseatEVE}, we have
\begin{eqnarray}
\mathbb{P}_{\mbox{\tiny ZF}}(\mathrm{E}|{\bf G})&\leq&n_t\mathbb{P}_{w \samp \mathcal{N}_{\sigma^2_{\mathrm{E}}}}\left(|w| < 1/2\right)\nonumber\\
&\leq&n_t\mathbb{P}_{w \samp \mathcal{N}_{1}}\left(|w| < |\sigma_{n_t}({\bf G})|/(2 m \alpha) \right).~\label{Pe:AE}
\end{eqnarray}
By comparing \eqref{Pe:AB} and \eqref{Pe:AE}, we see that the noise conditions for decoding ${\bf x}$ by users $\mathrm{B}$ and $\mathrm{E}$ are the same if both users have the same number of receive antennas $n_r'=n_r$ and the distributions of channels ${\bf G}$ and ${\bf H}$ are the same. This implies that user $\mathrm{E}$ is able to decode under the same constraints/conditions as $\mathrm{B}$. Moreover, if $n_r'> n_r$, then the adversary $\mathrm{E}$ is capable of decoding higher noise.	
\subsection{Asymptotic Probability of Error for Adversary}\label{SubSection:AsymptoticProbabilityofErrorforAdversary}
Before starting this section, we mention a Theorem from~\cite{Edelman89} regarding the least/largest singular value of matrix variate Gaussian distribution. This theorem relates the least/largest singular value of a Gaussian matrix to the number of its columns and rows asymptotically.
\begin{theorem}[\cite{Edelman89}]~\label{th:sv}
Let ${\bf M}$ be an $s\times t$ matrix with i.i.d. entries distributed as $\mathcal{N}_1$. If $s$ and $t$ tend to infinity in such a way that $s/t$ tends to a limit $y\in[1,\infty]$, then
\begin{equation}~\label{eq:advratio}
\sigma^2_t({\bf M})/s\rightarrow \left(1-\sqrt{1/y}\right)^2
\end{equation}
and
\begin{equation}~\label{eq:upLsv}
\sigma^2_1({\bf M})/s\rightarrow \left(1+\sqrt{1/y}\right)^2,
\end{equation}
almost surely.
\end{theorem}
We now analyze the asymptotic probability of error for eavesdropper using a ZF linear receiver.
\begin{theorem}~\label{th:PeZF}
Fix any real $\varepsilon,\varepsilon'>0$, and $y' \in [1,\infty]$, and suppose that $n_r'/n_t\rightarrow y'$ as $n_t \rightarrow \infty$. Then, for all sufficiently large $n_t$, the probability $\mathbb{P}_{\mbox{\tiny ZF}}(\mathrm{E})$ that $\mathrm{E}$ incorrectly decodes the message ${\bf x}$ using a ZF decoder is upper bounded by $\varepsilon$, if
\begin{equation}~\label{eq:PeZF}
m^2\alpha^2\leq \frac{n_r'\left((1-\sqrt{1/y'})^2-\varepsilon'\right)}{8\log(2n_t/\varepsilon)}.
\end{equation}
\end{theorem}
\begin{IEEEproof}
Let $\mathcal{G}$ be the set of all channel matrices ${\bf G}$ such that $\sigma_{n_t}^2({\bf G})\geq n_r'\left((1-\sqrt{1/y'})^2-\varepsilon'\right)$. Note that ${\bf G} \not\in \mathcal{G}$ with vanishing probability $o(1)$ as $n_t \rightarrow \infty$, by Theorem~\ref{th:sv}. We have:
\begin{eqnarray}
\mathbb{P}_{\mbox{\tiny ZF}}(\mathrm{E})\!\!\!\!\!\!&=&\!\!\!\!\!\mathbb{P}_{\mbox{\tiny ZF}}(\mathrm{E}|{\bf G}\in\mathcal{G})\mathbb{P}\left({\bf G}\in\mathcal{G}\right)\!+\!\mathbb{P}_{\mbox{\tiny ZF}}(\mathrm{E}|{\bf G}\notin\mathcal{G})\mathbb{P}\left({\bf G}\notin\mathcal{G}\right)\nonumber\\
\!\!\!\!\!&\leq&\!\!\!\! \mathbb{P}_{\mbox{\tiny ZF}}(\mathrm{E}|{\bf G}\in\mathcal{G})+\mathbb{P}\left({\bf G}\notin\mathcal{G}\right)\nonumber\\
\!\!\!\!\!&\leq&\!\!\!\! n_t\mathbb{P}_{w \samp \mathcal{N}_{1}}\left(|w| < |\sigma_{n_t}({\bf G})|/(2 m \alpha) \right)+o(1)\nonumber\\
\!\!\!\!\!&\leq&\!\!\!\! n_t\exp\left(-\sigma^2_{n_t}({\bf G})/\left(8m^2\alpha^2\right)\right)+o(1)\nonumber\\
\!\!\!\!\!&\leq&\!\!\!\! n_t\exp\left(\frac{-n_r'((1-\sqrt{1/y'})^2-\varepsilon')}{8m^2\alpha^2}\right)+o(1),\nonumber
\end{eqnarray}
where the first inequality is due the facts that $\mathbb{P}\left({\bf G}\in\mathcal{G}\right)\leq1$ and $\mathbb{P}_{\mbox{\tiny ZF}}(\mathrm{E}|{\bf G}\notin\mathcal{G})\mathbb{P}\left({\bf G}\notin\mathcal{G}\right)\leq\mathbb{P}\left({\bf G}\notin\mathcal{G}\right)$, the second inequality is true based on \eqref{Pe:AE} and Theorem~\ref{th:sv}, the third inequality uses the well-known upper bound $\exp\left(-x^2/2\right)$ for the tail of a Gaussian distribution and the last inequality follows from the definition of $\mathcal{G}$. By letting $\mathbb{P}_{\mbox{\tiny ZF}}(\mathrm{E}) \leq \varepsilon$, the sufficient condition \eqref{eq:PeZF} can be obtained.
\end{IEEEproof}
Comparing conditions \eqref{eq:const1} and \eqref{eq:PeZF}, we conclude that if $y'$ exceeds a small factor at most logarithmic in $n_t$, i.e. $y' = \mathcal{O}(\log n_t)$ we can have both conditions satisfied and yet Theorem~\ref{th:PeZF} shows that \aln{MIMO-Search} can be efficiently solved, i.e. this contradicts the hardness of the \aln{MIMO -Search} problem conjectured in~\cite{DG13} to hold for much larger polynomial ratios $y' = O(\poly(n_t))$. 

To analytically investigate the advantage of decoding at $\mathrm{B}$ over $\mathrm{E}$, we define the following advantage ratio.
\begin{definition}~\label{definition:advratio}
For fixed channel matrices ${\bf H}$ and ${\bf G}$, the ratio
\begin{equation}~\label{def:advratio}
\mbox{adv}\triangleq \sigma^2_{n_t}({\bf H})/\sigma^2_{n_t}({\bf G}),
\end{equation}
is called the advantage of $\mathrm{B}$ over $\mathrm{E}$.
\end{definition}
We note from \eqref{Pe:AB} and \eqref{Pe:AE} that $\mbox{adv}$ is the ratio between the maximum noise power tolerated by $\mathrm{B}$'s ZF decoder to the maximum noise power tolerated by $\mathrm{E}$'s ZF decoder, for the same decoding error probability in both cases. First, we study this advantage ratio asymptotically. We use Theorem~\ref{th:sv} to obtain the following result.
\begin{proposition}~\label{prop:rectangularMMPLCdisadv}
Let ${\bf H}_{n_r \times n_t}$ be the channel between $\mathrm{A}$ and $\mathrm{B}$ and ${\bf G}_{n_r' \times n_t}$ be the channel between $\mathrm{A}$ and $\mathrm{E}$, both with i.i.d. elements each with distribution $\mathcal{N}_1$. Fix real $y,y' \in [1,\infty]$, and suppose that $n_r/n_t\rightarrow y$ and $n_r'/n_t\rightarrow y'$ as $n_t \rightarrow \infty$. Then, using a \aln{SVD} precoding technique in \aln{MM-PLC}, we have $\mbox{adv} \rightarrow \left(\sqrt{y}-1\right)^2/\left(\sqrt{y'}-1\right)^2$ almost surely as $n_t \rightarrow \infty$.
\end{proposition}
\begin{IEEEproof}
Based on Theorem~\ref{th:sv} for ${\bf H}$ and ${\bf G}$, we have
$$\left\{\begin{array}{l}
\sigma^2_{n_t}({\bf H})/n_r\rightarrow (1-\sqrt{1/y})^2\\
\sigma^2_{n_t}({\bf G})/n_r'\rightarrow (1-\sqrt{1/y'})^2.
\end{array}\right.$$
Substituting the above two limits into \eqref{def:advratio} and using $n_r/n_r' = (n_r/n_t)/(n'_r/n_t) \rightarrow y/y'$, the result follows. 
\end{IEEEproof}
Note that $\mbox{adv}\rightarrow1$ is obtained in the case that $y=y'$ , which is equivalent to $n_r/n_r'\rightarrow1$. On the other hand $\mbox{adv}\rightarrow0$, if $y'/y = \infty$ which is equivalent to $n_r'/ n_r \rightarrow \infty$.
\subsection{General Precoding Scheme}\label{SubSection:GeneralPrecodingScheme}
One may wonder whether a different precoding method (again, assumed known to $\mathrm{E}$) than used above may provide a better advantage ratio for $\mathrm{B}$ over $\mathrm{E}$. 
Suppose that instead of sending $\tilde{\bf x} = {\bf V}{\bf x}$, user $\mathrm{A}$ precodes $\tilde{\bf x} = {\bf P}({\bf H}){\bf x}$, where ${\bf P} = {\bf P}({\bf H})$ is some other precoding matrix that depends on the channel matrix ${\bf H}$. Then, given the channel matrices, the analysis given in Section~\ref{Section:Zero-ForcingAttack} shows that using ZF decoding, $\mathrm{B}$'s decoding error probability will be bounded as $ n_t \exp(-\sigma^2_{n_t}({\bf H}{\bf P})/(8m^2\alpha^2))$, while $\mathrm{E}$'s decoding error probability will be bounded as $n_t \exp(-\sigma^2_{n_t}({\bf G}{\bf P})/(8m^2\alpha^2))$. Therefore, in this general case, the advantage ratio of maximum noise power decodable by $\mathrm{B}$ to that decodable by $\mathrm{E}$ at a given error probability generalizes from \eqref{def:advratio} to
\begin{equation}~\label{def:advratiogen}
\mbox{adv}\triangleq \sigma^2_{n_t}({\bf H}{\bf P})/\sigma^2_{n_t}({\bf G}{\bf P}).
\end{equation}
We now give an upper bound on the advantage ratio \eqref{def:advratiogen}. Let us first define
$$\mbox{advup}\triangleq\frac{\sigma^2_1({\bf H})}{\sigma^2_{n_t}({\bf G})}.$$
\begin{proposition}~\label{prop:rectangularMMPLCGeneralPrecoderdisadv}
Let ${\bf H}$ and ${\bf G}$ be as in Proposition~\ref{prop:rectangularMMPLCdisadv}. Then we have $\mbox{adv}\leq\mbox{advup}$. Furthermore, fix real $y,y' \in [1,\infty]$, and suppose that $n_r/n_t\rightarrow y$ and $n_r'/n_t\rightarrow y'$ as $n_t \rightarrow \infty$, so that $n_r'/n_r \rightarrow y'/y \triangleq \rho'$. Then, using a general precoding matrix ${\bf P}({\bf H})$ in \aln{MM-PLC}, we have
$\mbox{advup} \rightarrow \left(\sqrt{y}+1\right)/\left(\sqrt{y'}-1\right)^2$ almost surely as $n_t \rightarrow \infty$. Hence, in the case $n'_r=n_r$ and $y'=y \rightarrow \infty$, we have $\mbox{advup} \rightarrow 1$. Moreover, if $\mbox{advup} \rightarrow c$ for some $c \geq 1$, then $\min(y',\rho') \leq 9$. 
\end{proposition}
\begin{IEEEproof}
It is easy to see the two inequalities below hold for every ${\bf H}$, ${\bf G}$, and ${\bf P}$:
$$\left\{\begin{array}{l}
\sigma_{n_t}({\bf H}{\bf P}) \leq \sigma_1({\bf H})\sigma_{n_t}({\bf P}),\label{eq:ublsvproduct}\\
\sigma_{n_t}({\bf G}{\bf P}) \geq \sigma_{n_t}({\bf G})\sigma_{n_t}({\bf P}).\label{eq:lblsvproduct}
\end{array}
\right.$$
Hence, the advantage ratio \eqref{def:advratiogen} can be upper bounded as
\begin{equation}~\label{eq:advratioGeneralPrecoderMM-PLC}
\mbox{adv} \leq \frac{\sigma^2_1({\bf H})\sigma^2_{n_t}({\bf P})}{\sigma^2_{n_t}({\bf G})\sigma^2_{n_t}({\bf P})}=\frac{\sigma^2_1({\bf H})}{\sigma^2_{n_t}({\bf G})}=\mbox{advup}.
\end{equation}
Using Theorem~\ref{th:sv} for the the numerator and the denominator of the RHS  of \eqref{eq:advratioGeneralPrecoderMM-PLC}, respectively, and $n_r/n_r' \rightarrow y/y'$, we get
$$\mbox{advup} \rightarrow \frac{y(1+\sqrt{1/y})^2}{y'(1-\sqrt{1/y'})^2} = \left(\frac{\sqrt{y}+1}{\sqrt{y'}-1}\right)^2.$$
In the case $n_r'=n_r$ and $y=y'\rightarrow \infty$, the latter inequality gives $\mbox{advup} \rightarrow 1$. Also, the inequality $\left(\sqrt{y}+1\right)/\left(\sqrt{y'}-1\right)^2 \geq 1$ implies (using $y=y'/\rho'$) that $\rho' \leq 1/(1-2/\sqrt{y'})^2$, and the RHS of the latter is $\leq 9$ for all $y' \geq 9$, which implies $\min(y',\rho') \leq 9$.
\end{IEEEproof}
\section{Achievable Upper Bound on Advantage Ratio}~\label{AnUpperBoundonAdvantageRatio}
\begin{figure}[htb]%
  \begin{center}%
\includegraphics[width=8.5cm]{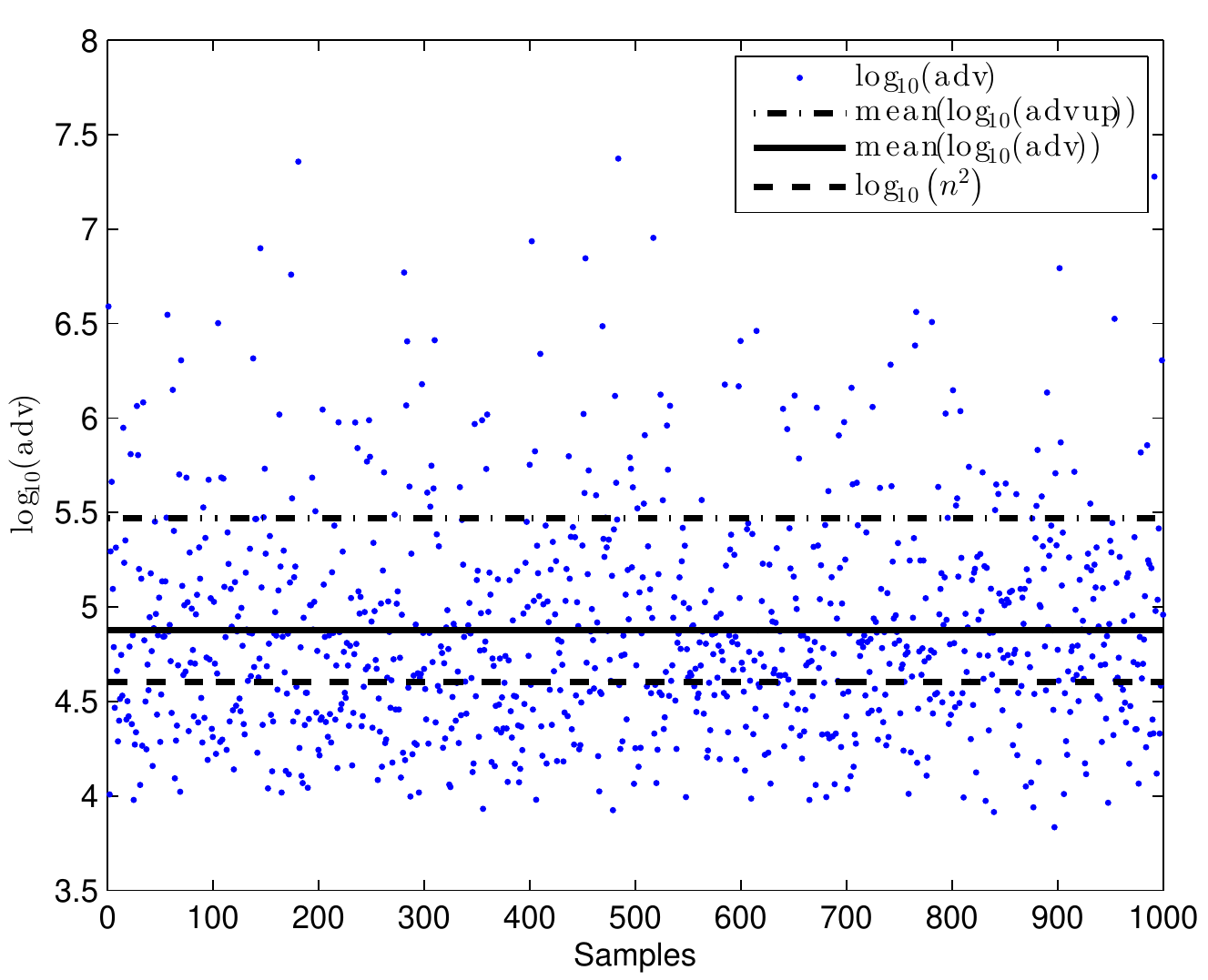}\vspace{-0.2cm}~\caption{\label{fig:advratioinverse} The advantage ratio \eqref{def:advratio} for $1000$ square channels of size $n=200$ using inverse precoder.}
  \end{center}
\vspace{-0.8cm}
\end{figure}

The above analysis shows that one cannot hope to achieve an advantage ratio greater than 1, if the the adversary uses a number of antennas significantly larger than used by the legitimate parties (by more than a constant factor). We now explore what advantage ratio can achieve if we add a new constraint to \aln{MM-PLC}, namely the number of adversary antennas is limited to be the same as the number of legitimate transmit and receive antennas. 
That is, we study the advantage ratio when the channel matrices ${\bf H}$ and ${\bf G}$ are square matrices and not rectangular. 
We show that under this simple constraint $n=n_t=n_r=n_r'$, the advantage ratio is capable of getting larger than $1$ and as big as $\mathcal{O}\left(n^2\right)$. We employ the following result in our analysis.
\begin{theorem}[\cite{Edelman89}]~\label{th:squarelsv}
Let ${\bf M}$ be a $t\times t$ matrix with i.i.d. entries distributed as $\mathcal{N}_1$. The least singular value of ${\bf M}$ satisfies
\begin{equation}~\label{eq:squarelsv}
\lim_{t\rightarrow\infty}\mathbb{P}\left[\sqrt{t}\sigma_t({\bf M})\geq x\right] = \exp\left(-x^2/2-x\right).
\end{equation}
\end{theorem}
We note that for a similar result on the largest singular value for square matrices, Theorem~\ref{th:sv} is enough. Using the above Theorem along with Theorem~\ref{th:sv}, one can further upper bound and estimate the advantage ratio. More precisely, we have
\begin{eqnarray}
\mbox{adv} &\leq& \sigma^2_1({\bf H})/\sigma^2_{n}({\bf G})\label{eq:in-1}\\
&\rightarrow &4n/\sigma^2_{n}({\bf G})=4n^2/\left(n\sigma^2_{n}({\bf G})\right),\label{eq:in0}
\end{eqnarray}
where \eqref{eq:in-1} is obtained based on \eqref{eq:advratioGeneralPrecoderMM-PLC}. As $n\rightarrow\infty$, based on Theorem~\ref{th:squarelsv}, the denominator of the RHS of \eqref{eq:in0} is $\mathcal{O}(1)$ except with probability $\leq \varepsilon$ for any fixed $\varepsilon>0$, and thus $\mbox{adv}$ is $\mathcal{O}\left(n^2\right)$ with the same probability. The following proposition is now outstanding.
\begin{proposition}~\label{prop:squareub}
Let $\varepsilon>0$ be fixed, ${\bf H}$ and ${\bf G}$ be $n\times n$ matrices as in Proposition~\ref{prop:rectangularMMPLCdisadv} with $n=n_t=n_r=n_r'$. Using a general precoder ${\bf P}({\bf H})$ to send the plain text ${\bf x}$, the maximum possible $\mbox{adv}$ that $\mathrm{B}$ can achieve over $\mathrm{E}$, is of order $\mathcal{O}\left(n^2\right)$, except with probability $\leq \varepsilon$.
\end{proposition}
The above proposition implies that user $\mathrm{B}$ \emph{may} be able to decode the message ${\bf x}$, with noise power up to $n^2$ times greater than $\mathrm{E}$ is able to handle. Such an advantage was not available in \aln{MM-PLC} scheme proposed in~\cite{DG13} due to the lack of constraint on the number of receive antennas for $\mathrm{E}$ and the use of SVD precoder. We present below experimental evidence that this upper bound can be approached using an \emph{inverse} precoder ${\bf P}({\bf H}) = {\bf H}^{-1}$. This inverse precoder may not be power efficient as it may need a lot of power enhancement at $\mathrm{A}$, however it gives us a benchmark on the achievable advantage ratio. In this framework, the equivalent channel between legitimate users is the identity matrix and the channel between users $\mathrm{A}$ and $\mathrm{E}$ is ${\bf G}{\bf H}^{-1}$. In Fig.~\ref{fig:advratioinverse}, we have shown the value of $\log_{10}\left(\mbox{adv}\right)$ for $1000$ square channel matrices of size $n=200$. For refrence, we also plot the mean value along with $\log_{10}\left(200^2\right)$. 
Clearly, in most cases the advantage ratio \eqref{def:advratio} is within a small factor (compared to $n^2$) of $n^2$.
\section{Summary and Directions for Future Work}~\label{sec7}
Our results suggest several natural open problems for future work. The implied contradiction between our first contribution and the conjectured hardness of \aln{MIMO-Search} in~\cite{DG13} for $n'_r/n_t = \mathcal{O}(\poly(n_t))$ implies either a polynomial-time algorithm for worst-case $\aln{GapSVP}_{\poly(n_t)}$ or that the complexity reduction of~\cite{DG13} (Theorem 1 of~\cite{DG13}) between \aln{MIMO-Search} and $\aln{GapSVP}_{\poly(n_t)}$ does not hold under the hardness condition of~\cite{DG13}. We believe the second possibility is the correct one, and that there is a gap in the proof of Theorem~1 of~\cite{DG13}. We do not yet know if the gap can be filled to give a worst-case to average-case reduction under a revised hardness condition. This is left for future work.

Our generalized upper bound on legitimate user to adversary \aln{ZF} decoding advantage suggests the complexity-based approach does not remove the needed linear limitation on the number of adversary antennas versus the number of legitimate party antennas, that is also suffered by previous information-theoretic methods. Can a more general complexity-based approach to physical-layer security avoid this limitation?

Finally, our positive result for the inverse precoder suggests that if the adversary is limited to have the same number of antennas as the legitimate parties, the complexity-based approach may provide practical security. This suggests the following questions: How secure is this inverse precoding scheme against more general decoding attacks (other than \aln{ZF})? Can a security reduction from a worst-case standard lattice problem be given for this case? How does the practicality of the resulting scheme compare to existing physical-layer security schemes based on information-theoretic security arguments? Can the efficiency of those schemes be improved by the complexity-based approach?
\vspace{-0.2cm}
%

\end{document}